\documentclass[fleqn,usenatbib]{mnras}

\usepackage[T1]{fontenc}

\DeclareRobustCommand{\VAN}[3]{#2}
\let\VANthebibliography\thebibliography
\def\thebibliography{\DeclareRobustCommand{\VAN}[3]{##3}\VANthebibliography}

\usepackage{graphicx}	
\usepackage{amsmath}	
\usepackage{amssymb}	
\usepackage{newtxtext,newtxmath}

\defcitealias{2021arXiv210209568V}{V\&B21}

\title[Dark Energy in Void-dominated Cosmology]{ Surface Tension of Cosmic Voids\\ as a Possible Source for Dark Energy}

\author[E. Yusofi et al.]{
E. Yusofi$^{1,2,3}$,\thanks{E-mail: e.yusofi@iauamol.ac.ir (Corresponding author)}
M. Khanpour$^{3,4}$,\thanks{E-mail: mtankhanpour@yahoo.com} B. Khanpour$^{5}$,,\thanks{E-mail: khanpourbh@gmail.com} M. A. Ramzanpour$^{1,3}$,,\thanks{E-mail: m.ramzanpour212@gmail.com} M. Mohsenzadeh$^{6}$,\thanks{E-mail: mohsenzadeh@qom-iau.ac.ir}
\\
$^{1}$Department of Physics, Faculty of Basic Sciences, Ayatollah Amoli Branch, Islamic Azad University, Amol, Iran\\$^{2}$School of Astronomy, Institute for Research in Fundamental Sciences(IPM), P. O. Box 19395-5531,Tehran, Iran\\$^{3}$Innovation and Management Research Center, Ayatollah Amoli Branch, Islamic Azad University, Amol, Mazandaran, Iran
\\$^{4}$Faculty of Basic Sciences, Ayatollah Amoli Branch, Islamic Azad University, Amol, Iran\\$^{5}$Department of Physics, Faculty of Basic Sciences, University of Mazandaran, P. O. Box 47416-95447, Babolsar, Iran\\$^{6}$Department of Physics, Faculty of Basic Sciences, Qom Branch, Islamic Azad University, Qom, Iran}

\date{Accepted XXX. Received YYY; in original form ZZZ}

\pubyear{2021}

\begin{document}
\label{firstpage}
\pagerange{\pageref{firstpage}--\pageref{lastpage}}
\maketitle

\begin{abstract}
The cosmological constant is estimated by considering the surface tension of supervoids in a void-dominated cosmic fluid by which we can get a possible source of dark energy. Looking at voids as bubbles, we define the concept of surface tension which is shown to have an almost constant value for supervoids that are enclosed by superclusters. The surface tensions of voids are computed by dimensional method for galaxies and superclusters with different values for each group. At large scale which vast voids are dominant the positive cosmological constants obtained  of order $ (\simeq+10^{-52} {\rm m^{-2}} )$, which are very close to those given by $Planck$.
\end{abstract}




\section{Introduction}
\par As we know the existence of a type of material fluid with negative pressure remains an unsolved puzzle in physics and cosmology, because the lowest perceived pressure for any type of ordinary matter fluid is zero. But one of the ways to justify the current accelerating expansion in the universe is to have negative pressure for the cosmic fluid at large scale ~\citep{SupernovaSearchTeam:2004lze, SupernovaCosmologyProject:1998vns, WMAP:2003elm, Planck:2015fie, Cheng:2005fn,Srivastava:2008grc} . 
\par In the ideal gas law it is assumed that the particles are point-like without any volumes and interactions, but in the present physical cosmic fluid, there exist some over-dense and under-dense regions that some phenomena occurring between cosmic objects, for example the merging of them is a common phenomenon at cosmological scale. Galaxies and galaxy clusters, and even voids are all always evolving and merging together to form larger ones. Although the volume of the merged galaxies are small fractions of the total cosmos volume, but taking into account the contribution of vast voids, the volumes of these voids become large fractions of the present cosmic web and their contributions will grow up ~\citep{Cowen2015VastCV, Pisani:2015jha,Hamaus:2016wka,Khanpour:2017das,Koren:2017GCV}. As voids expand, galaxies are squeezed in between them, and sheets
and filaments form the void boundaries. This view is supported by computer simulations and numerical studies ~\citep{WEYGAERT:2011csd,Padilla:2005ea,Martel:1990ae}. Also, novel cosmological constraints obtained from cosmic voids and exact calibrations demonstrates that can be expect the immense potential to use cosmic voids for cosmology in current and future data ~\citet{Hamaus:2020cbu}.
\par Scientists usually consider only matter (galaxies and their clusters) as the active part of the universe, so telescopes and probes focus on this part of the cosmic fluid. But the much larger under-dense part of the current universe, the supervoids, are considered as the ineffective part in the dynamics of the universe. In this article for the first time, we take the role of cosmic voids very seriously and consider the effect of theirs \textit{surface tension} on  global and local scale to solve important unresolved problems in physical cosmology. Several papers have pointed out that vast voids are not only a key element of the cosmic mass and volume distribution, but also one of the purest probes for global cosmic parameters~\citep{Bos:2012wq,Pisani:2015jha,Hamaus:2016wka}. We consider the current cosmic fluid as a mixture of two evolving and merging parts. These include the over-dense galactic part on the one hand and the under-dense voids part on the other. Given that voids make up a much larger contribution of the current cosmic fluid, we will show that this void-dominant fluid will create additional effective negative pressure (with a negative 'cosmic equation of state' i.e. $w_{\rm {eff}} < 0$), on the large scale. As an important consequence of this hypothesis, we will also be able to obtain possible estimates for the cosmological constant by dimensional calculating of the surface tensions on the supervoids boundary that enclosed by superclusters. Our hypothesis is related to the surface tension of supervoids, in which their average effect can create an effective negative pressure at cosmic scales.\\
Our main thesis of this paper is the surface tension of cosmic
voids due to the existence of inhomogeneities and hence their role in global expansion is closely related to the "backreaction" issue ~\citep{Rasanen:2011ki,Wiltshire:2011vy,Clarkson:2011zq,Buchert:2011sx,Buchert:2015iva}. By definition, backreaction  is the effect of inhomogeneities caused by matter and \textit{geometry} on the global expansion of the universe ~\citet{Buchert:2015iva}. In some papers, Green and Wald believe that backreaction of inhomogeneities is irrelevant in cosmology~\citep{Green:2010qy,Green:2013yua}. As we know, in Newtonian context only inhomogeneities of the matter part can be considered and the geometry part is considered flat. Newtonian limit of GR automatically implies that the geometry is Euclidean, in particular that the intrinsic curvature vanish everywhere~\citep{Buchert:2011sx,Buchert:2017obp}. But when we consider the global scale of the universe, the Newtonian framework - due to neglecting the role of curved geometry- can not provide the desired results, and the Buchert-Ehlers theorem results in a zero backreaction ~\citet{Buchert:1995fz}.\\
In a recent new work, the effect of inhomogeneities in the Newtonian framework with a non-Euclidean topology is investigated~\citet{Vigneron:2021tpi}. The effect of the backreaction in the context of general relativity have been widely studied~\citep{Buchert:1999er,Buchert:2001sa,Buchert:2019mvq}, in which both the role of matter and geometric inhomogeneities are considered on a global scale. So the consequence will be the non-zero backreaction, and hence it can affect the dynamics and the accelerating expansion of the universe. The model presented here can achieve this important request with a heuristic calculation of the surface tension from the void-cluster interface.\\
The model presented in this paper is in the context of relativistic cosmology. To achieve this goal we have already considered second-order terms in the \textit{cosmic equation of state} as
$$ P=w\rho+ b\rho^2, $$
that the details are given in~\citet{Khanpour:2017das}.\\
\textit{Why the role of supervoids and their surface tension is taken seriously in this article?}
When we consider significant the second term- as an interacting term- in the EoS of cosmic fluid; there must be physical objects that are interacting (or merging) with each other on the cosmic scale. Because the scales are so large, the best possible candidate for these merging objects are the largest ones \textit{i.e.} merging supervoids that are enclosed by superclusters~\citep{Sutter:2014kda,Cowen2015VastCV}. By taking into account the effect of the merging of vast voids on the cosmic web together with the enlargement of empty spaces over time at low redshift in N-body simulations~\citet{Adermann:2017izw}, as well as its similarity to the behavior of bubbles in hot overflowing milk~\citet{Yusofi_2010}; the idea of using bubble surface tension will be considered as a possible model for studying effects of supervoids pressure in accelerating expansion. Recently, similar to our idea, the surface tension hypothesis has also been proposed by Ortiz to explain the accelerated expansion of the universe and some other important challenges of physical cosmology~\citet{Ortiz:2020noa}. In his paper the cosmic system under study was considered homogeneous and isotropic, while the formation of cosmic structure and its inhomogeneities is ignored. But in our work, inhomogeneities in the structure of the universe are the main factor of the production of supervoids/superclusters and the resulting their surface tension.
\par So in the Sec. ~\ref{Sec 2.} we will introduce void-dominated cosmic fluid and by the drops-bubbles mixed fluid model show that in such a fluid we will have an effective negative pressure at large scale. In the Sec. ~\ref{Sec 3.} and ~\ref{Sec 4.}, by a dimensional calculating of the surface tension, we will obtain a possible estimation for the values of cosmological constants for two groups of cosmic objects $i.e.$ superclusters and galaxies. Some results, predictions and discussions on the model will be presented in final section.
\begin{figure}
	\centering
    \includegraphics[width=2.5 in]{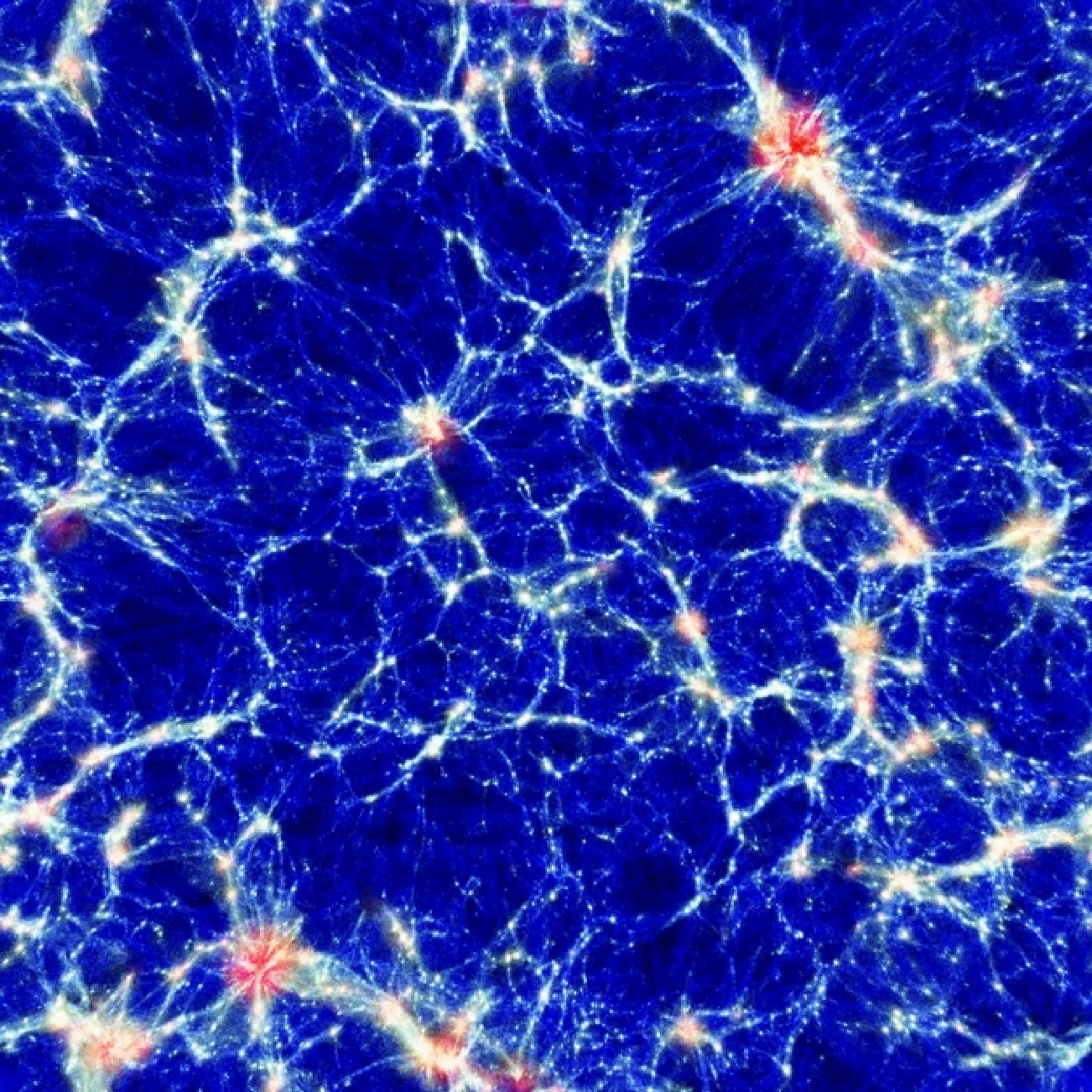}
	\caption{The cosmic web is mostly occupied by supervoids: The wight and red areas are superclusters and filaments of galaxies, blue areas are voids and wight dots are single galaxies inside the cosmic voids (https://sci.esa.int/web/planck/-/51104-numerical-simulation-of-the-cosmic-web; with permission).}
	\label{Fig1}
\end{figure}
\begin{figure}
	\centering
	\includegraphics[width=2.5 in]{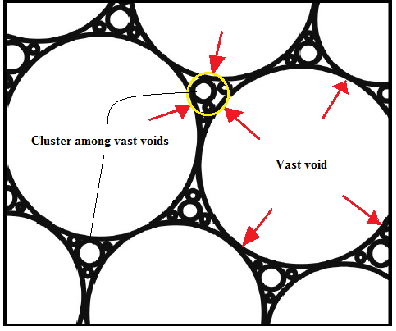}
	\caption{Schematic of void-dominated cosmic fluid that occupied by supervoids: Negative pressure with effective force of supervoids on galaxies in the walls (black circle) acts as the positive pressure with effective attractive  force on local scale (yellow circle)}.
	\label{Fig2}
\end{figure}
\section{Pressure and Surface Tension of Cosmic Voids}
\label{Sec 2.}
\par Let's consider the current cosmic web  in Fig. \ref{Fig1} containing a network of the voids, in which several clusters and filaments, small and vast voids are merging to each other (see the area bounded to the black rectangular in Fig. \ref{Fig2}). What is certain is that the universe is in the void-dominated state in the large scale overview. The continuous merging of voids causes the bubbles to grow larger, and this increasing in the bubble radius results in an extra effective  repulsive force on the particles (i.e., galaxies) on the surface of the bubbles (i.e., voids). Under such conditions, cosmic fluid is clustered like merging bubbles and is situated in the void-dominated phase. \\
In the model proposed in this paper, we imagine galaxies and their clusters as 'drops' and under dense spaces (voids) between them as 'bubbles'~\citet{Yusofi_2010}. In a mixture of water drops and bubbles, bubbles on their surface move the fine drops away from their center to make themselves larger, while large drops attract fine drops to their center to grow larger. If the void wall forms a set of galaxy clusters and superclusters, we will have a bubble whose difference in internal and external pressure comes from the Young-Laplace formula ~\citep{Butt:2003pci,Reichl:2016msp}
\begin{equation}\label{delp}
\Delta{P} = \frac{2\gamma}{\bar{r}}.
\end{equation}
Here, $\gamma$ represents the surface tension for drop (bubble) with average radius $\bar{r}$. 
The clusters of galaxies behave like drops, so their pressure on the test particles/galaxies is positive with an attractive gravity effect (yellow circle in Fig. \ref{Fig2}). These galaxies experience attractive force in the clusters-dominated areas at local scales, but at large scale that we have void-dominated areas, the situation is completely different. Galaxies that mostly accumulate on the surface of voids, would experience  effective negative pressure, because of the fact that voids inclosed by superclusters are expanding, \emph{and hence pushing the galaxies away from each other}. Therefore, inspired by (\ref{delp}), two factors are effective in the pressure of voids on cosmic scales. The first factor is the surface tension $\gamma$, due to the attraction between material particles that make up the disk-shaped objects on the surface of voids, and the second factor is due to the curvature ($\frac{1}{\bar{r}} $) of these cosmic voids . 
\par Because the net effective pressure of the cosmic web at large scale essentially dominated by supervoids and their pressure, in which the curvature term $({\frac{1}{\bar{r}}})$ takes a negative sign in this case  (see Fig. \ref{Fig2}). Here the negative sign means that the average pressure on galaxies from the expanding supervoids at cosmic scales acts in reverse to the pressure from the superclusters at local scales, and causing the galaxies to move away. Fortunately, the observational data confirm negative pressure that EoS parameter $w$ has a very narrow range around $w=-1$ with more likelihood to the side of $w\lesssim -1$ ~\citep{Srivastava:2008grc, Planck:2018vyg}.
\par In what follows, we will show that the negative pressure coming from surface tension and curvature of voids can justify the presence of dark energy at large scale, and probably dark matter in local scales.
\section{Dimensional Calculation of Surface Tension}
\label{Sec 3.}
\begin{figure}
	\centering
	\includegraphics[width=2.5 in]{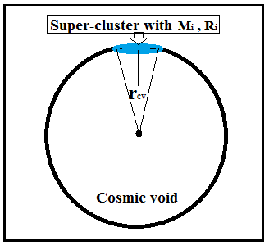}
	\caption{Schematic picture of supervoid with shell of superclusters on it surface.}
	\label{Fig3}
\end{figure}

The supervoids make up the main volume of the universe at cosmic scale. Therefore, the thickness of the walls is small compared to the large volume of the supervoids and they can almost be considered as an idealised surface. The presence of a separating surface between the under-dense and the over-dense areas creates a surface tension that produce pressure difference between them. However, the interior of walls is well observed and the effect of wall thickness could be investigated in a more accurate model.\\
Given the above ideal assumption for the surface tension of supervoids, if we consider the location of clusters on the surface of supervoids, as a result of this consideration, the mass and energy of galaxies and their clusters are distributed on the shells of voids. Therefore, the   \emph{energy-to-area} ratio i.e. surface tension $\gamma$ for the disc-shaped objects can be calculated by using the following,
\begin{equation}
\label{hard715}
\gamma_{i}\equiv\frac{Energy}{Area}={\frac{M_{i}c^2}{\pi R_{i}^2}}.
\end{equation}
In the above relation $M_i$ and $R_i$ are binding mass and mean radius of the cosmic objects such as galaxies, clusters and superclusters (Fig. \ref{Fig3}). For example, the amount of surface tension or surface energy for the Laniakea supercluster with $M_{3} = 1.0 \times 10^{47} {\rm kg}$ and $R_{3} = 2.4 \times 10^{24} {\rm m}$ is obtained as,
\begin{equation}
\label{hard716}
\gamma_{3} \approx 0.50 \times 10^{15} {\rm J.{m^{-2}}}.
\end{equation}
Because the surface tension is an intensive quantity ~\citep{Butt:2003pci,Zemansky:2011hat}, it is expected that the approximate values of surface energy for the superclusters at large scale are the same order ($\sim {\cal{O}} (10^{15}){\rm J.m^{-2}} $), but the energy density is not the same for these objects.

\section{An Estimation of the Cosmological Constant with Vast Void Pressure}
\label{Sec 4.}
In the standard cosmology with cosmological constant $\Lambda$, we have ~\citep{Cheng:2005fn, Srivastava:2008grc}
\begin{equation}
\label{hard81}
(\frac{\dot{a}}{a})^2={\frac{8\pi{G}}{3}}(\rho_{\rm {matter}})+\frac{\Lambda{c^2}}{3}-\frac{kc^2}{a^2}.
\end{equation}
Considering contribution of voids energy density $\rho_{\rm {void}}$, instead of cosmological constant in the void-dominated (or quasi-vauum dominated~\citet{Yusofi:2018lqb}) cosmic fluid, we can modify Friedmann’s equation to
\begin{equation}
\label{hard80}
(\frac{\dot{a}}{a})^2={\frac{8\pi{G}}{3}}(\rho_{\rm {matter}}+\rho_{\rm {void}})-\frac{kc^2}{a^2}.
\end{equation}

Since a void-dominated phase of cosmic fluid can be regarded as a quasi-vacuum dominated state i.e. $\rho_{\rm {void}} = \rho_{\rm {\Lambda}}$, therefore two relations $(\ref{hard80})$ and $(\ref{hard81})$ are equivalent and we can define
\begin{equation}
\label{hard82}
\Lambda=\frac{8\pi{G}\rho_{\rm {void}}}{c^2}.
\end{equation}
with consideration $\rho_{\rm {void}} = P_{\rm {void}}/w{c^2}$,
\begin{equation}
\label{hard83}
\Lambda=\frac{8\pi{G}{P_{\rm {void}}}}{w{c^4}}.
\end{equation}
After formation of the structure  in the cosmic scales, its dominant part is supervoids that are enclosed by superclusters. The effective pressure supposed in these scales stems from the surface tension produced on the interface between supervoids and superclusters. Therefore, due to the predominance of voids in large structure heterogeneities, the pressure difference in (\ref{delp}), can be equivalent to the effective pressure caused by the voids i.e. $\Delta{P} \simeq {P_{\rm {void}}}$, and we can write,
\begin{equation}
\label{hard838}
\Lambda =\frac{8\pi{G}}{{w{c^4}}}\frac{2\gamma}{\bar r_{\rm {void}}}.
\end{equation}
According to the relation obtained above, $\Lambda$ in addition to the surface tension $\gamma$, depends another factor i.e. $(\frac{1}{r})$, which is similar to the curvature term, we can choose both negative and positive sign for it depending on whether  under study is global or local. Therefore, we can expect a positive and negative cosmological constant for global and local scales, respectively. Therefore we note that according to the sign of curvature term one can talk of a de
Sitter world matter ($\Lambda_{\rm cs}$ positive, pressure negative) or an anti-de Sitter world matter ($\Lambda_{\rm ls}$ negative, pressure positive)~\citet{Gazeau:2021}.\\
Since the effective force on galaxies from surface tension of supervoid at global scale is repulsive (with negative effective pressure), for dark energy EoS $i.e. (w < 0)$, relations (\ref{hard715}), (\ref{hard838}) and (\ref{hard819}) lead to a cosmological constant with the positive sign as follows,
\begin{equation}
\label{hard84}
\Lambda_{\rm {cs}} =\frac{16{G}M_{\rm i}}{{w}\bar{r}_{\rm {cs}}{c^2}{R_{\rm i}}^2} > 0.
\end{equation}
\begin{table}
	\caption{Surface tension $\gamma_{\rm i}$, on the vast voids shell and $\Lambda_{\rm i}$ for disc shape cosmic objects}
	\label{tab:1}       
	\begin{center}
		\begin{tabular}[width=\textwidth]{llllll}
			
			\hline\noalign{\smallskip}
			i. Cosmic&\quad $M_{\rm i}$&\quad $R_{\rm i}$&\quad \quad $\gamma_{\rm i}$ & \quad \quad$\Lambda_{\rm i}$ \\
			\quad object&($10^{47}{\rm {kg}}$) & ($10^{24}{\rm {m}}$) & $(10^{15}{\rm {J.m^{-2}}})$ & $(10^{-52}{\rm {m^{-2}}})$ \\
			\noalign{\smallskip}\hline
			\noalign{\smallskip}\hline
			1. Corona Sc&\quad $0.20$ &\quad $1.50$ &\quad $0.25$ &\quad $0.6645$ \\
			2. Virgo Sc&\quad $0.03$ &\quad $0.50$ &\quad $0.34$ &\quad $0.8970$ \\
			3. Laniakea Sc &\quad $1.00$ &\quad $2.40$ &\quad $0.50$ &\quad $1.2979$\\
			4. Caelum Sc&\quad $4.00$ &\quad $ 4.30$ &\quad $0.62$ &\quad $1.6172$\\
			\noalign{\smallskip}\hline\noalign{\smallskip}
			5. Milky Way &\quad $0.00002$ &\quad $ 0.00087$ &\quad $75.74$ &\quad $1975.32$\\
			6. Andromeda &\quad $0.00003$ &\quad $ 0.001$ &\quad $85.98$ &\quad $2242.68$\\
			7. UGC02885 &\quad $0.000004$ &\quad $ 0.00036$ &\quad $88.46$ &\quad $2307.28$\\
			\noalign{\smallskip}\hline
			\noalign{\smallskip}\hline
		\end{tabular}
	\end{center}
\end{table}
According to a recent study, an intergalactic void is responsible for pushing the galaxies such as Andromeda or Milky Way at increasing speed through the universe ~\citet{Hoffman:2017ako}. Cosmic voids typically have a diameter of 100 ${\rm {Mpc}}$ i.e. $ \bar{r}_{\rm void} = 1.54 \times 10^ {+24} {\rm {m}} $. Also, $G = 6.67 \times 10^{-11} {\rm {m^{3}.kg^{-1}.s^{-2}}}$ , $ w_{\rm {Planck}} =-1.03 \pm 0.03 $, and $c = 3\times 10^ {8}  {\rm {m.s^{-1}}} $. By putting $(\ref{hard716})$ and these values in equation $(\ref{hard84})$, we can estimate the following value for the cosmological constant for $\gamma_3 = 0.50 \times 10^{15}{\rm {J.m^{-2}}}$,
\begin{equation}
\label{hard85}
\Lambda_3 =  1.2979 \times  10^{-52} {\rm {m^{-2}}}.
\end{equation}
By using of the $Planck$ 2018 values of $\Omega_{\Lambda} = 0.6889\pm 0.0056 $ and $H_0 = 67.66\pm 0.42 {\rm (km/s)/{Mpc}}$, $\Lambda$ has the value of ~\citet{Planck:2018vyg}
\begin{equation}
\label{hard86}
\Lambda_{\rm {obs}}= 3\Omega_{\Lambda}\left(\frac{H_0}{c}\right)^2  = 1.1056 \times 10^{-52} {\rm {m^{-2}}}.
\end{equation}
The cosmological constant value in relation $(\ref{hard84})$ depends on both the surface energy and the radius of different voids, but for a mean size of the cosmic voids, the order obtained for cosmological constant at cosmic scale $\Lambda_{\rm {cs}}$, in each case is very close to the observational value $(\ref{hard86})$ and according to the first four rows of Table 1. is in the following range 
\begin{equation}
\label{hard876}
0.6645\times 10^{-52} {\rm {m^{-2}}}\leq  \Lambda_{{\rm {cs}}} \leq 1.6172\times 10^{-52} {\rm {m^{-2}}}. 
\end{equation}
Fortunately, the observational value $(\ref{hard86})$ for cosmological constant is in this range.\\
Looking at Table 1, we will find that smaller surface energy and cosmological constant are obtained for larger cosmic objects. At large scale due to the presence of galactic superclusters on the one hand and supervoids on the other, we will obtain the least value of surface tension. Therefore, the cosmological constant value, for example for Laniakea supercluster is equal to (\ref{hard85}), which is very close to the observational value for the cosmological constant (\ref{hard86}). 
\par In Table 1, by considering the mass and approximate radius of the disc-shaped cosmic objects, we could obtain the surface tension for the two groups of objects.  Then we will get the cosmological constant corresponding to each surface tensions according to (\ref{hard838}). The values obtained in Table 1 show that for larger objects we will have smaller surface tension and therefore less cosmological constant and vice versa. The lowest surface tension is related to the superclusters that make up the walls of large cosmic voids, but the largest surface tension is related to the galaxies that make up the walls of local voids. So correspondingly, the maximum amount of cosmological constant for galaxies is $ \Lambda_{\rm {max}} = 2307.2800 \times 10^{-52} {\rm {m^{-2}}}$ approximately $3472$ times that of the minimum of it ($\Lambda_{\rm {min}} = 0.6645 \times 10^{-52} {\rm {m^{-2}}})$.
\par As a consequence of the dominance of supervoids with negative curvature at cosmic scale,  we have shown by an interesting heuristic line of reasoning, that almost constant surface energy for the superclusters on the boundaries of supervoids acts as a cosmological constant and consequently acts as a driver in the accelerating motion of galaxies at large scales.
\par Many studies have shown that the presence of inhomogeneities in the structure of the universe plays a role in global expansion~\citep{Rasanen:2011ki,Wiltshire:2011vy,Clarkson:2011zq,Buchert:2011sx,Buchert:2015iva}. So more precisely in our model superclusters and supervoids as the largest present inhomogeneities are the result of structure formation. These two largest objects in the universe can coexist and can produce effective pressure in global and local scales. So far we have shown that the effective pressure caused by supervoids on a global scale acts as a potential source of dark energy. On the other hand, we speculate that collective force from merging supervoids may act as a possible source for dark matter. On this basis, we also predict that there are places in the universe that can not have dark matter. We will address this issue in the near future work. 
\section{Summary and Some Possible Consequences of the Model}
We have proposed that the present cosmic fluid composed of two dynamical parts. On the one hand matters are clustering and on the other hand voids are merging, simultaneously. Since the dominant volume of the present cosmic web is formed by merging voids, based the our model, the negative pressure or repulsive gravity is inevitable at large scale. That is, the cosmic gas is considered as mixture of so many voids (bubbles), each of supervoids has repulsive force on galaxies, and hence the total effective pressure of the universe at large scale would become negative or repulsive. In other words, voids merge with other voids to form larger voids. As a result of this process, the galaxies on the shell of supervoids move away from each other. As an interesting point, the merging of the cosmic voids at their surfaces may produce both additional contracting (positive pressure) at local over-dense scales and accelerating (negative pressure) at large under-dense ones, simultaneously. 
\par By dimensional calculating of the surface energy on the boundary of voids, we have obtained almost the same value of surface tensions for the supervoids enclosed by superclusters from the order of $ {\cal{O}} (10^{15}) {{\rm J.m^{-2}}}$. Also, we have calculated acceptable range for global cosmological constant values that the observational value for $\Lambda$ is in this range.\\ As we know, the scale of the largest supervoids grows in the standard model and in the proposed model with cosmological time, and this probably has an effect on the  cosmic equation-of-state and effective pressure of cosmic fluid. The effective equation-of-state for the cosmic fluid in our model has a negative value which is of the order of $"1"$. Since, our calculations are dimensional and approximate, so small changes in the amount of $w_{\rm eff}$ do not have much effect on our calculations and results. In addition, it should be noted that in the real model the voids are very non-spherical and our results are at most a rough order-of-magnitude argument.\\
The key point of this work is that the universe has evolved from a matter-dominated cosmic fluid to a \emph{void-dominated} one (instead of a vacuum-dominated). Therefore, the presence of such supervoids in this period is considered as the possible driver of the universe expansion and creates additional pressure on ordinary matter.\\ 
Highlights include:\\
(1) The present cosmic fluid consists of two coexistent evolving parts, namely merging superclusters and merging supervoids.\\
(2) In the proposed model,  we have considered a fundamental role for cosmic voids and their surface tensions in the dynamics of the present universe at global scale.\\
(3) Both negative effective pressure at the global scale and positive ones at local scales are introduced.\\
(4) The almost constant surface tension on the supervoids shells reproduces a cosmological constant whose magnitude is as same as that confirmed by observational $Planck$ data.
\par Since, the cosmological constant value depends on surface energy of clusters and the radius of voids, i.e. $\Lambda \sim \gamma/r_{{\rm {void}}}$ , at large-scale which supervoids are dominant, we have obtained very small values for the cosmological constants for the model of order $ (+10^{-52} {{\rm m^{-2}}})$, which are very close to those given by $Planck$ 2018. However, on local scales we were led to the larger values with negative sign $ (\Lambda_{\rm {ls}}\simeq -10^{-49} {{\rm m^{-2}}})$. It can be speculated that the surface tensions of supervoids on a local scale provide an additional effective force that holds galaxies, clusters, and superclusters together. Finally, the validity of the proposed model can also be further examined in the three years of observed dark energy survey data (DES Y3)\citet{DES:2021wwk} as well as the Euclid mission~\citet{Amendola:2016saw}. As a final point, we also expect that the proposed model is capable to solve the Hubble tension. Attempts are being made to investigate this issue.

 \section*{Data availability}
 The data underlying this article will be shared on reasonable request to the corresponding author.
\section*{Acknowledgements}
The authors would like to thank the anonymous reviewer for his very useful comments and suggestions for improving the quality of the article. EY would like to acknowledge Dr. A. Talebian for his constructive discussions and for his help. This work has been supported by the Islamic Azad University, Ayatollah Amoli Branch, Amol, Iran.

\bibliographystyle{mnras}
\bibliography{zmy3-MNRAS} 

\begin{thebibliography}{}
\makeatletter
\relax
\def\mn@urlcharsother{\let\do\@makeother \do\$\do\&\do\#\do\^\do\_\do\%\do\~}
\def\mn@doi{\begingroup\mn@urlcharsother \@ifnextchar [ {\mn@doi@}
  {\mn@doi@[]}}
\def\mn@doi@[#1]#2{\def\@tempa{#1}\ifx\@tempa\@empty \href
  {http://dx.doi.org/#2} {doi:#2}\else \href {http://dx.doi.org/#2} {#1}\fi
  \endgroup}
\def\mn@eprint#1#2{\mn@eprint@#1:#2::\@nil}
\def\mn@eprint@arXiv#1{\href {http://arxiv.org/abs/#1} {{\tt arXiv:#1}}}
\def\mn@eprint@dblp#1{\href {http://dblp.uni-trier.de/rec/bibtex/#1.xml}
  {dblp:#1}}
\def\mn@eprint@#1:#2:#3:#4\@nil{\def\@tempa {#1}\def\@tempb {#2}\def\@tempc
  {#3}\ifx \@tempc \@empty \let \@tempc \@tempb \let \@tempb \@tempa \fi \ifx
  \@tempb \@empty \def\@tempb {arXiv}\fi \@ifundefined
  {mn@eprint@\@tempb}{\@tempb:\@tempc}{\expandafter \expandafter \csname
  mn@eprint@\@tempb\endcsname \expandafter{\@tempc}}}

\bibitem[\protect\citeauthoryear{Abbott et~al.}{Abbott
  et~al.}{2021}]{DES:2021wwk}
Abbott T. M.~C.,  et~al., 2021

\bibitem[\protect\citeauthoryear{Ade et~al.}{Ade et~al.}{2016}]{Planck:2015fie}
Ade P. A.~R.,  et~al., 2016, \mn@doi [Astron. Astrophys.]
  {10.1051/0004-6361/201525830}, 594, A13

\bibitem[\protect\citeauthoryear{Adermann, Elahi, Lewis  \& Power}{Adermann
  et~al.}{2017}]{Adermann:2017izw}
Adermann E.,  Elahi P.~J.,  Lewis G.~F.,   Power C.,  2017, \mn@doi [Mon. Not.
  Roy. Astron. Soc.] {10.1093/mnras/stx657}, 468, 3381

\bibitem[\protect\citeauthoryear{Aghanim et~al.}{Aghanim
  et~al.}{2020}]{Planck:2018vyg}
Aghanim N.,  et~al., 2020, \mn@doi [Astron. Astrophys.]
  {10.1051/0004-6361/201833910}, 641, A6

\bibitem[\protect\citeauthoryear{Amendola et~al.}{Amendola
  et~al.}{2018}]{Amendola:2016saw}
Amendola L.,  et~al., 2018, \mn@doi [Living Rev. Rel.]
  {10.1007/s41114-017-0010-3}, 21, 2

\bibitem[\protect\citeauthoryear{Bos, van~de Weygaert, Dolag  \& Pettorino}{Bos
  et~al.}{2012}]{Bos:2012wq}
Bos E. G.~P.,  van~de Weygaert R.,  Dolag K.,   Pettorino V.,  2012, \mn@doi
  [Mon. Not. Roy. Astron. Soc.] {10.1111/j.1365-2966.2012.21478.x}, 426, 440

\bibitem[\protect\citeauthoryear{Buchert}{Buchert}{2000}]{Buchert:1999er}
Buchert T.,  2000, \mn@doi [Gen. Rel. Grav.] {10.1023/A:1001800617177}, 32, 105

\bibitem[\protect\citeauthoryear{Buchert}{Buchert}{2001}]{Buchert:2001sa}
Buchert T.,  2001, \mn@doi [Gen. Rel. Grav.] {10.1023/A:1012061725841}, 33,
  1381

\bibitem[\protect\citeauthoryear{Buchert}{Buchert}{2018}]{Buchert:2017obp}
Buchert T.,  2018, \mn@doi [Mon. Not. Roy. Astron. Soc.]
  {10.1093/mnrasl/slx160}, 473, L46

\bibitem[\protect\citeauthoryear{Buchert \& Ehlers}{Buchert \&
  Ehlers}{1997}]{Buchert:1995fz}
Buchert T.,  Ehlers J.,  1997, Astron. Astrophys., 320, 1

\bibitem[\protect\citeauthoryear{Buchert \& R\"as\"anen}{Buchert \&
  R\"as\"anen}{2012}]{Buchert:2011sx}
Buchert T.,  R\"as\"anen S.,  2012, \mn@doi [Ann. Rev. Nucl. Part. Sci.]
  {10.1146/annurev.nucl.012809.104435}, 62, 57

\bibitem[\protect\citeauthoryear{Buchert et~al.}{Buchert
  et~al.}{2015}]{Buchert:2015iva}
Buchert T.,  et~al., 2015, \mn@doi [Class. Quant. Grav.]
  {10.1088/0264-9381/32/21/215021}, 32, 215021

\bibitem[\protect\citeauthoryear{Buchert, Mourier  \& Roy}{Buchert
  et~al.}{2020}]{Buchert:2019mvq}
Buchert T.,  Mourier P.,   Roy X.,  2020, \mn@doi [Gen. Rel. Grav.]
  {10.1007/s10714-020-02670-6}, 52, 27

\bibitem[\protect\citeauthoryear{Butt, Graf  \& Kappl}{Butt
  et~al.}{2003}]{Butt:2003pci}
Butt H.~J.,  Graf K.,   Kappl M.,  2003, Physics and Chemistry of Interfaces.
WILEY-VCH, Germany

\bibitem[\protect\citeauthoryear{Cheng}{Cheng}{2010}]{Cheng:2005fn}
Cheng T.~P.,  2010, {Relativity, gravitation, and cosmology: A basic
  introduction}.
Oxford Univ. Pr., Oxford, UK

\bibitem[\protect\citeauthoryear{Clarkson, Ellis, Larena  \& Umeh}{Clarkson
  et~al.}{2011}]{Clarkson:2011zq}
Clarkson C.,  Ellis G.,  Larena J.,   Umeh O.,  2011, \mn@doi [Rept. Prog.
  Phys.] {10.1088/0034-4885/74/11/112901}, 74, 112901

\bibitem[\protect\citeauthoryear{Cohen-Tannoudji \& Gazeau}{Cohen-Tannoudji \&
  Gazeau}{2021}]{Gazeau:2021}
Cohen-Tannoudji G.,  Gazeau J.~P.,  2021, ] {10.20944/preprints202105.0320.v2}

\bibitem[\protect\citeauthoryear{Cowen}{Cowen}{2015}]{Cowen2015VastCV}
Cowen R.,  2015, Nature

\bibitem[\protect\citeauthoryear{Green \& Wald}{Green \&
  Wald}{2011}]{Green:2010qy}
Green S.~R.,  Wald R.~M.,  2011, \mn@doi [Phys. Rev. D]
  {10.1103/PhysRevD.83.084020}, 83, 084020

\bibitem[\protect\citeauthoryear{Green \& Wald}{Green \&
  Wald}{2013}]{Green:2013yua}
Green S.~R.,  Wald R.~M.,  2013, \mn@doi [Phys. Rev. D]
  {10.1103/PhysRevD.87.124037}, 87, 124037

\bibitem[\protect\citeauthoryear{Hamaus, Pisani, Sutter, Lavaux, Escoffier,
  Wandelt  \& Weller}{Hamaus et~al.}{2016}]{Hamaus:2016wka}
Hamaus N.,  Pisani A.,  Sutter P.~M.,  Lavaux G.,  Escoffier S.,  Wandelt
  B.~D.,   Weller J.,  2016, \mn@doi [Phys. Rev. Lett.]
  {10.1103/PhysRevLett.117.091302}, 117, 091302

\bibitem[\protect\citeauthoryear{Hamaus, Pisani, Choi, Lavaux, Wandelt  \&
  Weller}{Hamaus et~al.}{2020}]{Hamaus:2020cbu}
Hamaus N.,  Pisani A.,  Choi J.-A.,  Lavaux G.,  Wandelt B.~D.,   Weller J.,
  2020, \mn@doi [JCAP] {10.1088/1475-7516/2020/12/023}, 12, 023

\bibitem[\protect\citeauthoryear{Hoffman, Pomarede, Brent~Tully  \&
  Courtois}{Hoffman et~al.}{2017}]{Hoffman:2017ako}
Hoffman Y.,  Pomarede D.,  Brent~Tully R.,   Courtois H.,  2017, ]
  {10.1038/s41550-016-0036}

\bibitem[\protect\citeauthoryear{Khanpour, Yusofi  \& Khanpour}{Khanpour
  et~al.}{2017}]{Khanpour:2017das}
Khanpour M.,  Yusofi E.,   Khanpour B.,  2017

\bibitem[\protect\citeauthoryear{Koren}{Koren}{2017}]{Koren:2017GCV}
Koren M.,  2017, Scince

\bibitem[\protect\citeauthoryear{Martel \& Wasserman}{Martel \&
  Wasserman}{1990}]{Martel:1990ae}
Martel H.,  Wasserman I.,  1990, \mn@doi [Astrophys. J.] {10.1086/168208}, 348,
  1

\bibitem[\protect\citeauthoryear{Ortiz}{Ortiz}{2020}]{Ortiz:2020noa}
Ortiz C.,  2020, \mn@doi [Int. J. Mod. Phys. D] {10.1142/S0218271820501151},
  29, 2050115

\bibitem[\protect\citeauthoryear{Padilla, Ceccarelli  \& Lambas}{Padilla
  et~al.}{2005}]{Padilla:2005ea}
Padilla N.~D.,  Ceccarelli L.,   Lambas D.~G.,  2005, \mn@doi [Mon. Not. Roy.
  Astron. Soc.] {10.1111/j.1365-2966.2005.09500.x}, 363, 977

\bibitem[\protect\citeauthoryear{Perlmutter et~al.}{Perlmutter
  et~al.}{1999}]{SupernovaCosmologyProject:1998vns}
Perlmutter S.,  et~al., 1999, \mn@doi [Astrophys. J.] {10.1086/307221}, 517,
  565

\bibitem[\protect\citeauthoryear{Pisani, Sutter, Hamaus, Alizadeh, Biswas,
  Wandelt  \& Hirata}{Pisani et~al.}{2015}]{Pisani:2015jha}
Pisani A.,  Sutter P.~M.,  Hamaus N.,  Alizadeh E.,  Biswas R.,  Wandelt B.~D.,
    Hirata C.~M.,  2015, \mn@doi [Phys. Rev. D] {10.1103/PhysRevD.92.083531},
  92, 083531

\bibitem[\protect\citeauthoryear{R\"as\"anen}{R\"as\"anen}{2011}]{Rasanen:2011ki}
R\"as\"anen S.,  2011, \mn@doi [Class. Quant. Grav.]
  {10.1088/0264-9381/28/16/164008}, 28, 164008

\bibitem[\protect\citeauthoryear{Reichl}{Reichl}{2016}]{Reichl:2016msp}
Reichl L.~E.,  2016, A Modern Course in Statistical Physics,.
WILEY-VCH, Germany

\bibitem[\protect\citeauthoryear{Riess et~al.}{Riess
  et~al.}{2004}]{SupernovaSearchTeam:2004lze}
Riess A.~G.,  et~al., 2004, \mn@doi [Astrophys. J.] {10.1086/383612}, 607, 665

\bibitem[\protect\citeauthoryear{Spergel et~al.}{Spergel
  et~al.}{2003}]{WMAP:2003elm}
Spergel D.~N.,  et~al., 2003, \mn@doi [Astrophys. J. Suppl.] {10.1086/377226},
  148, 175

\bibitem[\protect\citeauthoryear{Srivastava}{Srivastava}{2008}]{Srivastava:2008grc}
Srivastava S.~K.,  2008, General Relativity and Cosmology.
PHI, India

\bibitem[\protect\citeauthoryear{Sutter, Elahi, Falck, Onions, Hamaus, Knebe,
  Srisawat  \& Schneider}{Sutter et~al.}{2014}]{Sutter:2014kda}
Sutter P.~M.,  Elahi P.,  Falck B.,  Onions J.,  Hamaus N.,  Knebe A.,
  Srisawat C.,   Schneider A.,  2014, \mn@doi [Mon. Not. Roy. Astron. Soc.]
  {10.1093/mnras/stu1845}, 445, 1235

\bibitem[\protect\citeauthoryear{Van De~Weygaert \& Platen}{Van De~Weygaert \&
  Platen}{2011}]{WEYGAERT:2011csd}
Van De~Weygaert R.,  Platen E.,  2011, \mn@doi [International Journal of Modern
  Physics: Conference Series] {10.1142/s2010194511000092}, 01, 41–66

\bibitem[\protect\citeauthoryear{Vigneron}{Vigneron}{2021}]{Vigneron:2021tpi}
Vigneron Q.,  2021

\bibitem[\protect\citeauthoryear{Wiltshire}{Wiltshire}{2011}]{Wiltshire:2011vy}
Wiltshire D.~L.,  2011, \mn@doi [Class. Quant. Grav.]
  {10.1088/0264-9381/28/16/164006}, 28, 164006

\bibitem[\protect\citeauthoryear{Yusofi}{Yusofi}{2018}]{Yusofi:2018lqb}
Yusofi E.,  2018, \mn@doi [Mod. Phys. Lett. A] {10.1142/S021773231850219X}, 33,
  1850219

\bibitem[\protect\citeauthoryear{Yusofi \& Mohsenzadeh}{Yusofi \&
  Mohsenzadeh}{2010}]{Yusofi_2010}
Yusofi E.,  Mohsenzadeh M.,  2010, \mn@doi [Astronomy Education Review]
  {10.3847/aer2009035}, 9

\bibitem[\protect\citeauthoryear{Zemansky}{Zemansky}{2011}]{Zemansky:2011hat}
Zemansky M.,  2011, Heat And Thermodynamics.
McGraw-Hill Education, India

\makeatother
\end{thebibliography}
\end{document}